\documentclass[aps,prl,reprint,superscriptaddress]{revtex4-1}

\usepackage{amssymb}
\usepackage{amsmath}    
\usepackage{graphicx}   
\usepackage{graphics}
\usepackage{verbatim}   
\usepackage{color}      
\usepackage{subfigure}  
\usepackage{hyperref}   
\begin{document}


\title{Volume and porosity thermal regulation in lipid mesophases by coupling mobile ligands to soft membranes}


\author{Lucia Parolini}
\affiliation{Biological and Soft Systems Sector, Cavendish Laboratory, University of Cambridge, CB3 0HE, Cambridge, United Kingdom}
\author{Bortolo M. Mognetti}
\affiliation{Interdisciplinary Center for Nonlinear Phenomena and Complex Systems \&
Service de Physique des Syst\'{e}mes Complexes et M\'{e}canique Statistique, Universit\'{e} libre de Bruxelles, Campus Plaine, CP 231, Blvd du Triomphe, B-1050 Brussels, Belgium.}
\author{Jurij Kotar}
\affiliation{Biological and Soft Systems Sector, Cavendish Laboratory, University of Cambridge, CB3 0HE, Cambridge, United Kingdom}
\author{Erika Eiser}
\affiliation{Optoelectronics Group, Cavendish Laboratory, University of Cambridge, CB3 0HE, Cambridge, United Kingdom}
\author{Pietro Cicuta}
\affiliation{Biological and Soft Systems Sector, Cavendish Laboratory, University of Cambridge, CB3 0HE, Cambridge, United Kingdom}
\author{Lorenzo Di Michele}
\email[]{ld389@cam.ac.uk}
\affiliation{Biological and Soft Systems Sector, Cavendish Laboratory, University of Cambridge, CB3 0HE, Cambridge, United Kingdom}


\date{\today}

\begin{abstract}
Short DNA linkers are increasingly being exploited for driving specific self-assembly of Brownian objects. DNA-functionalised colloids can assemble into ordered or amorphous materials with tailored morphology. Recently, the same approach has been applied to compliant units, including emulsion droplets and lipid vesicles.  The liquid structure of these substrates introduces new degrees of freedom: the tethers can diffuse and rearrange, radically changing the physics of the interactions. Unlike droplets, vesicles are extremely deformable and DNA-mediated adhesion causes significant shape adjustments. We investigate experimentally the thermal response of pairs and networks of DNA-tethered liposomes and observe two intriguing and possibly useful collective properties: negative thermal expansion and tuneable porosity of the liposome networks. A model providing a thorough understanding of this unexpected phenomenon is developed, explaining the emergent properties out of the interplay between the temperature-dependent deformability of the vesicles and the DNA-mediated adhesive forces.
\end{abstract}

\pacs{}

\maketitle


\section{Introduction}
The seminal work of Mirkin~\cite{Mirkin_Nature_1996} and Alivisatos~\cite{Alivisatos_Nature_1996} introduced the idea of exploiting Watson-Crick base-pairing interaction~\cite{Watson-Crick_Nature_1953} to drive the self-assembly of Brownian objects. Together with the introduction of all-DNA nanostructures~\cite{Chen_Nature_1991}, this concept revolutionised our perspective on nucleic acids, which today are viewed as the prototype of programmable materials.\\
Thanks to the selectivity and thermal reversibility of the hybridisation interaction, DNA-functionalised nano- and colloidal particles~\cite{Di-Michele_PCCP_2013} can be made to assemble into ordered~\cite{Nykypanchuk_Nature_2008,Mirkin_Nature_2008,Mirkin_Science_2011,Park_NMat_2010,Hill_NanoLetters_2008} or amorphous materials~\cite{Varrato_PNAS_2012,Di-Michele_NatComm_2013,Di-Michele_SoftMatter_2014}  with programmable morphology.\\ 
To date, the highly controllable structure of hybrid DNA materials is not matched by a comparably tailorable response to external stimuli. {The few exceptions are limited to complex melting/aggregation behaviours controlled by suitably designed DNA coatings that allow step-wise activation of the interactions, either in response to temperature changes~\cite{Di-Michele_NatComm_2013}, or aided by external fields and photo-activated permanent linkers \cite{Leunissen_SoftMatter_2009}. Melting can also be controlled through competing linkages formed within the particles~\cite{Angioletti-Uberti_NMat_2012,Mognetti_SoftMatter_2012}, or with DNA strands dispersed in solution~\cite{McGinley_SoftMatter_2013,Eze_SoftMatter_2013,Baker_SoftMatter_2013,Tison_SoftMatter_2010}. Beyond melting/aggregation, structural responsiveness to external stimuli has only been achieved for nanoparticle aggregates, where competing linkers added in solution are able to significantly change the length of DNA bonds and thereby the density of the aggregates~\cite{Maye_NatNano_2010}.} A more versatile response could boost the applicability of hybrid DNA materials beyond the current biomedical applications to gene regulation~\cite{Rosi_Science_2006,Wu_JACS_2014} and molecular diagnostics~\cite{Cutler_JACS_2012}.\\
Greater flexibility could be achieved by replacing solid building blocks with more compliant units based on self-assembled phospholipid structures. These include oil-in-water emulsion droplets~\cite{Pontani_PNAS_2012,Feng_SoftMatter_2013,Hadorn_PNAS_2012}, liposomes~\cite{Beales_JPCA_2007,Beales_SoftMatter_2011,Hadorn_Langmuir_2013,Hadorn_PLOSOne_2010,Beales_ACIS_2014}, and hybrid substrates obtained by covering solid colloidal particles with a lipid bilayer~\cite{Meulen_JACS_2013}. On these liquid interfaces, DNA tethers can freely diffuse and rearrange unpon binding, introducing completely new physical effects that influence the interactions. Unlike droplets and lipid-coated particles, liposomes are extremely deformable and DNA-mediated adhesion causes significant shape adjustments.\\
In this article, we experimentally investigate the coupling between DNA-mediated adhesion and mechanical deformation of giant liposomes. We observe a striking response to temperature changes leading to negative thermal expansion and tuneable porosity of the liposome networks. These counterintuitive effects emerge from the interplay between the temperature-dependent mechanical properties of the liposomes and the mobility of the DNA linkers, and cannot be replicated with stiffer materials, including solid particles and oil droplets. The latter, because of the high interfacial tension as compared to the strength of DNA-mediated adhesion, are nearly undeformable. We focus on the quantitative understanding of the morphological changes and interactions in pairs of identical liposomes and rationalise our experimental observations by means of a detailed model that, with a single fitting parameter, is capable of predicting the temperature dependence of all the morphological observables as well as the fraction of formed DNA bonds. The success of our theoretical description confirms the key role played by confinement and combinatorial entropic effects on the interactions mediated by multiple -- mobile ligands, and unveils the mechanisms by which they are coupled to the morphology of the substrates. These findings can have important implications in understanding biological ligand-receptor adhesion.
We envisage potential applications of responsive DNA-lipid mesophases as tuneable sieves in the micrometer scale and smaller, drug delivery materials, mechanically and rheologically controllable scaffolds for cell regeneration, food material, or cosmetics.\\


\section{Results}
\subsection{Negative thermal expansion in DNA-GUVs}
\begin{figure}[ht!]
\includegraphics[width=8.5cm]{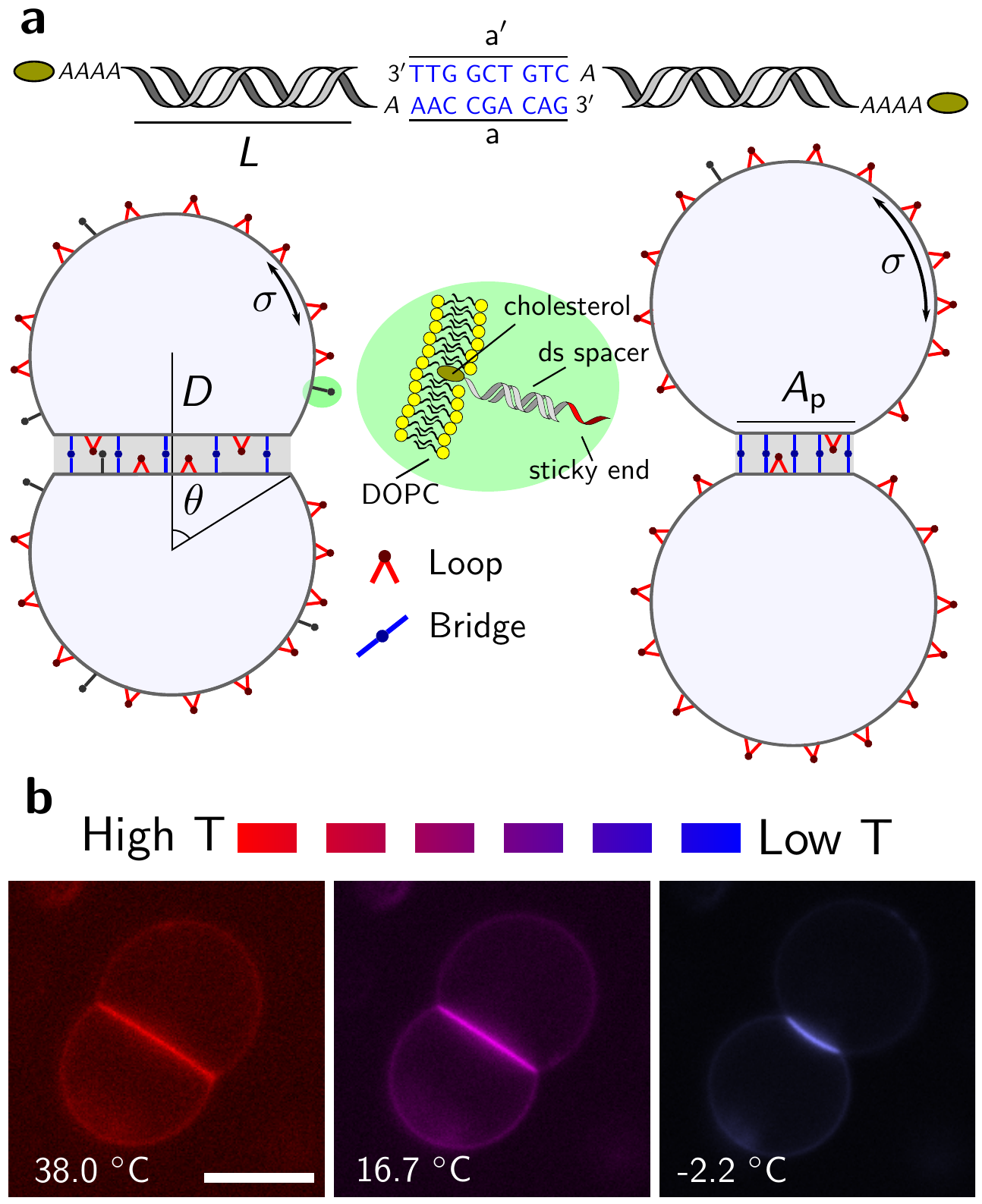}%
\caption{\label{Figure1}\textbf{Thermal response in pairs of DNA-linked vesicles.} \textbf{a}~Not-to-scale schematic view of a pair of adhering DNA-GUVs and of two bound tethers. The sticky ends $a$ and $a'$ are marked in blue, the dangling base $A$ and the flexible spacer $AAAA$ in italic font. \textbf{b}~Colour-coded epifluorescence images of a pair of adhering DNA-GUVs at decreasing temperatures (from left to right). Scale bar, 10~$\mu$m. For a full sequence see {Supplementary Movie} 1.}
\end{figure}

Giant unilamellar vesicles (GUVs) of 1,2-Dioleoyl-sn-glycero-3-phosphocholine (DOPC) are prepared via electroformation~\cite{Angelova_FD_1989,Angelova_PCPS_1992}. The liposomes are then functionalised with hydrophobically modified, pre-assembled, DNA constructs. These are made of a 43-base-pair long double stranded spacer with a cholesterol anchor at one end and an unpaired single-stranded DNA (ssDNA) sequence, a sticky end, at the other end~(Fig.~\ref{Figure1}a). To improve flexibility, an unpaired adenine base is left between the sticky end and the spacer; four unpaired bases are left between the spacer and the cholesterol. The anchoring takes place through the insertion of the  cholesterol into the bilayer's hydrophobic core~\cite{Beales_JPCA_2007,Beales_SoftMatter_2011,Hadorn_Langmuir_2013,Hadorn_PLOSOne_2010,Beales_ACIS_2014, Meulen_JACS_2013}. Anchored DNA can freely diffuse on the bilayer surface~\cite{Meulen_JACS_2013}. Half of the sticky ends carry a DNA sequence $a$, the second half carries the complementary sequence $a'$. Sticky ends $a$ bind to $a'$, forming either intra-vesicle loops or inter-vesicle bridges~\cite{Leunissen_NMat_2009}. The latter are responsible for the attractive force between adhering vesicles. The DNA can be visualised optically through fluorescent staining by means of an intercalating dye. Full details on the experimental protocol are provided in the Methods and Supplementary Methods sections.\\


In Fig.~\ref{Figure1}b we show epifluorecence microscopy snapshots and a schematic view of a pair of adhering vesicles. The DNA-mediated adhesive forces cause the deformation of the GUVs and the appearance of a quasi-flat adhesion patch. Bridges are confined to this area, causing a local increase in the overall DNA concentration and consequently in the fluorescence intensity (see also {Supplementary Movie} 1). The fluorescent emission from the free (non-adhering) portions of the membranes is due to both loops and unbound tethers, present both outside and within the patch.\\
At temperatures $\approx 40^\circ$C the patch takes up a considerable portion of the membranes, with contact angles $\theta$ as large as $60^\circ$. Upon cooling we observe the emergence of an unexpected phenomenon: the patch  becomes brighter, indicating an increase in DNA concentration, and shrinks (Fig.~\ref{Figure1}b). We quantify this effect by reconstructing the shape of pairs of adhering vesicles from equatorial cross-sections captured by epifluorescence microscopy, as illustrated in Fig.~\ref{Figure2}a. Full details on the imaging and image analysis protocols are provided in the Methods section.\\
In Fig.~\ref{Figure2}b-d we show the temperature dependence of the contact angles $\theta_{1/2}$, the patch area $A_\mathrm{p}$, and the distance $D$ measured between the centres of a typical pair of DNA-GUVs.  We focus our attention on pairs of vesicles having similar, but never identical, size: labels 1 and 2 refer to quantities measured for the two vesicles in the pair. Both $\theta_{1/2}$ and $A_\mathrm{p}$ decrease monotonically upon cooling, dropping to a tiny fraction of their initial values when $T \lesssim 5^\circ$C. Strikingly, the bond distance $D$ increases by a factor $\gtrsim1.5$, leading to a negative thermal expansion along the direction of the bond between the vesicles.
In Fig.~\ref{Figure2}e-f we show the radii $R_{1/2}$ of the spherical section of the vesicles and the overall surface areas $A_{1/2}$, both of which decrease as the temperature is decreased. The overall volumes $V_{1/2}$, shown in Fig.~\ref{Figure2}g, are nearly constant or display a slight decrease.\\
The shrinkage of the adhesion patch upon cooling appears at first inconsistent with the well-understood strengthening of DNA-mediated interactions~\cite{Di-Michele_PCCP_2013,Angioletti-Uberti_NMat_2012,Dreyfus_PRL_2009}, which would instead lead to a larger adhesion region. This apparent paradox can be at first rationalised by considering the temperature-dependent mechanical response of the GUVs.
Let us consider an isolated vesicle and introduce the reduced volume~\cite{Tordeux_PRE_2002,Ramachandran_Langmuir_2010}
\begin{equation}\label{eqn:reducedvolume}
v=\frac{\frac{3V}{4\pi}}{\left(\frac{A}{4\pi}\right)^{3/2}},
\end{equation}
where $V$ is the inner volume and $A$ the surface area of the vesicle. Isolated liposomes with $v=1$ are perfect spheres, with membrane tension $\sigma=0$. This condition occurs if $V=V_0=4\pi R_0^3 /3$ and $A=A_0=4\pi R_0^2$, where $R_0$ is a reference radius. However, the surface area of lipid bilayers expands significantly with temperature. Around a reference temperature $T_\mathrm{0}$ the area $A$ of the GUV can be expressed as
\begin{equation}\label{eqn:unstretched}
A=\tilde{A}=A_0\left[1+\alpha(T-T_0)\right],
\end{equation} 
where $\alpha$ is the area thermal expansion coefficient~\cite{Evans_JPC_1987}. The thermal expansion of the inner water solution is comparatively negligible, therefore, if we consider water impermeable vesicles, we can assume constant volume $V=V_0$ and rewrite the temperature-dependent reduced volume as
\begin{equation}
v=\left[1+\alpha(T-T_0)\right]^{-3/2}.
\end{equation}
For $T>T_0$, that is $v<1$, vesicles have an \emph{excess area}, resembling deflated balloons. In this regime, GUVs are easily deformable and any attractive force capable of suppressing thermal fluctuations will cause the adhesion of neighbouring vesicles. The resulting contact angle $\tilde{\theta}$ can be easily calculated from geometry for the case of two identical vesicles (see Methods section) and decreases monotonically with temperature, which explains the experimental trend (dashed line in Fig.~\ref{Figure2}b). For $T<T_0$ isolated GUVs become turgid spheres with $v>1$ and $\sigma>0$. In this regime $\tilde{\theta}=0$.\\

\subsection{Modelling membrane deformation and DNA-mediated adhesion}
The qualitative explanation provided above does not describe the role played by the DNA ligands. For a complete understanding of the emergent response we consider the interaction free energy between a pair of identical DNA-GUVs
\begin{equation}\label{eqn:energy0}
U=U_\mathrm{membrane}+U_\mathrm{DNA}+U_0.
\end{equation}
The first term on the right-hand side of Eq. \ref{eqn:energy0} accounts for the repulsive contributions arising from membrane deformation. The term $U_\mathrm{DNA}$ represents the free energy of the mobile linkers and encodes the adhesive forces as well as other more subtle effects, investigated in detail below. The free energy $U_0$, calculated for two isolated DNA-GUVs,  is taken as a reference.\\ Three effects contribute to $U_\mathrm{membrane}$: the stretching of the bilayer, its bending, and the suppression of thermal fluctuations caused by the adhesion to the second membrane. The last two contributions can be neglected in the limit of strong adhesion, as demonstrated by Ramachandran \emph{et al.}~\cite{Ramachandran_Langmuir_2010}. The sphericity of the non-adhering portions of the DNA-GUVs confirms the dominant role played by membrane tension and the little effect of bending elasticity. The interaction free energy can thus be rewritten to include only the stretching term
\begin{equation}
U(\hat{\theta})=K_\mathrm{a}\frac{\left[A(\hat{\theta})-\tilde{A}\right]^2}{\tilde{A}}+U_\mathrm{DNA}(\hat{\theta})+U_0\label{eqn:energy1},
\end{equation}
where $A$ is the overall area of a vesicle, $\tilde{A}$ is the temperature-dependent unstretched area described by Eq. \ref{eqn:unstretched}, and $K_\mathrm{a}$ is the elastic stretching modulus. All the geometrical observables ($A$, $A_\mathrm{p}$, $D$, $R$, and $V$), and therefore the interaction energy of our system, can be parametrised using the contact angle as the only independent variable, as demonstrated in the Methods section. When indicating the independent variable, the contact angle is labeled as $\hat{\theta}$ to distinguish it from its equilibrium counterpart $\theta$.\\
Let us now focus on the calculation of $U_\mathrm{DNA}$.
We start by providing an expression for the hybridisation free-energy between a pair of tethered DNA-linkers, forming either bridges (b) or loops (l)~\cite{Mognetti_SoftMatter_2012}
\begin{equation}\label{eqn:singleenergy}
\Delta G_\mathrm{b/l}(\hat{\theta})=\Delta G^0 - T \Delta S^\mathrm{conf}_\mathrm{b/l}(\hat{\theta}).
\end{equation}
The term $\Delta G^0=\Delta H^0-T\Delta S^0$ indicates the hybridisation free energy for the untethered constructs. Enthalpic and entropic contributions can be estimated using the conventional nearest-neighbours model~\cite{SantaLucia_PNAS_1998}, corrected for the presence of inert tails that, as discussed below, cause a shift in $\Delta G^0$ due to electrostatic effects~\cite{Di-Michele_JACS_2014}.\\
The entropic contribution $\Delta S_\mathrm{b/l}^\mathrm{conf}(\hat{\theta})$ is characteristic of tethered DNA linkers and quantifies the loss of configurational freedom following the formation of a bond~\cite{Mognetti_SoftMatter_2012,Dreyfus_PRL_2009}. As sketched in Fig.~\ref{Figure3}a, we can model the DNA tethers as freely pivoting rigid rods of length $L$ with freely diffusing tethering points. Sticky ends are modelled as point particles. Assuming a uniform distance between the membranes of adhering GUVs equal to $L$, we can calculate
\begin{eqnarray}
\Delta S_\mathrm{b/l}^\mathrm{conf}(\hat{\theta})&=&k_\mathrm{B} \log \left[\frac{1}{\rho_0 L} \frac{\hat{A}_\mathrm{b/l}(\hat{\theta})}{A(\hat{\theta})}\right]\nonumber \\
&=&\Delta S^\mathrm{rot} +\Delta S_\mathrm{b/l}^\mathrm{trans}(\hat{\theta})\label{eqn:confent}.
\end{eqnarray}
We can identify two contributions to the entropic loss. The term $\Delta S^\mathrm{rot}$ accounts for the hinderance in the pivoting motion and can be estimated as~\cite{Leunissen_JCP_2011}
\begin{equation}
\Delta S^\mathrm{rot} =k_\mathrm{B}\log\left[\frac{1}{4 \pi \rho_0 L^3}\right]\label{eqn:rotent},
\end{equation}
where $\rho_0$=1~M $=0.6$ nm$^{-3}$ is a reference concentration. By quantifying $L=14.5$ nm as the length of the double stranded spacer~\cite{Smith_Science_1996}, the expression in Eq. \ref{eqn:rotent} predicts a repulsion $-T\Delta S^\mathrm{rot}=10.0$ $k_\mathrm{B}T$, independent of the bond type (loop or bridge) and of the geometry of the GUVs (see Fig.~\ref{Figure3}d).\\
The term $\Delta S^\mathrm{trans}_\mathrm{b/l}$ accounts for the loss of translational entropy following the formation of a bond, and is unique to systems of mobile tethers. It can be estimated as
\begin{equation}
\Delta S^\mathrm{trans}_\mathrm{b/l}(\hat{\theta})=k_\mathrm{B}\log\left[\frac{4\pi L^2 \hat{A}_\mathrm{b/l}(\hat{\theta})}{A^2(\hat{\theta})}\right]\label{eqn:transent},
\end{equation}
where $\hat{A}_\mathrm{b/l}$ indicates the area of the region over which loops and bridges can freely diffuse. For loops, $\hat{A}_\mathrm{l}$ is equal to the total area of the vesicle $A$. For bridges $\hat{A}_\mathrm{b}=A_\mathrm{p}^\mathrm{e}=A_\mathrm{p}+2\pi L R$, equal to the patch area augmented by the narrow ring-shaped region within which bridge formation is permitted (Fig.~\ref{Figure3}b-c and Methods section).  Since $A_\mathrm{p}^\mathrm{e}<A$, the contribution $\Delta S^\mathrm{trans}_\mathrm{b/l}$ particularly hinders bridge formation. For a typical pair of vesicles $-T\Delta S^\mathrm{trans}_\mathrm{l}\approx13$ $k_\mathrm{B}T$, while $-T\Delta S^\mathrm{trans}_\mathrm{b}\approx16-20$ $k_\mathrm{B}T$, with a clear increase at low $\hat{\theta}$ due the shrinking of the patch area (Fig.~\ref{Figure3}d).
Details on the derivation of Eq.~\ref{eqn:confent} are provided in the Methods section.\\
Equations \ref{eqn:singleenergy}-\ref{eqn:transent} contain an expression for the hybridisation $\Delta G_\mathrm{b/l}$ for the formation of a single bridge or loop. Note that coupling between $\Delta G_\mathrm{b/l}$ and the geometry of the GUVs (i.e. $\hat{\theta}$) occurs only via the translational entropy term.\\

For a fixed geometry, that is fixed $A(\hat{\theta})$, $A_\mathrm{p}(\hat{\theta})$ and consequently $\Delta G_\mathrm{b/l}(\hat{\theta})$, the overall hybridisation free energy $U_\mathrm{hyb}$ between a pair of vesicles can be estimated by analytically evaluating the equilibrium free energy of a system of $2N$ linkers with negligible steric interactions. As derived in the Methods section we obtain~\cite{Varilly_JChemPhys_2012,Angioletti-Uberti_JCP_2013}
\begin{equation}\label{eqn:DNAenergy1}
U_\mathrm{hyb}(\hat{\theta}) = k_\mathrm{B}TN\left[4 \log \left(1 - x_\mathrm{b} - x_\mathrm{l}\right) + 2 x_\mathrm{b} + 2 x_\mathrm{l}\right],
\end{equation}
where
\begin{equation}\label{eqn:DNAenergy2}
x_\mathrm{b/l}(\hat{\theta}) = \frac{q_\mathrm{b/l} \left(1+2q_\mathrm{l} + 2 q_\mathrm{b} - \sqrt{1 + 4q_\mathrm{l} + 4q_\mathrm{b} } \right) }{ 2 \left(q_\mathrm{l} + q_\mathrm{b}\right)^2 }
\end{equation}
are the equilibrium fractions of linkers involved in bridges/loops, and
\begin{equation}\label{eqn:DNAenergy3}
q_\mathrm{b/l}(\hat{\theta})=\exp \left(-\beta \Delta G_\mathrm{b/l}^*\right).\\
\end{equation}
In Eq.~\ref{eqn:DNAenergy3} we define $\Delta G_\mathrm{b/l}^*=\Delta G_\mathrm{b/l}-k_\mathrm{B} T \log N$. The non-extensive combinatorial term $-k_\mathrm{B} T \log N $ is once again unique of systems of mobile linkers and accounts for the fact that any tether can potentially bind to $N$ partners. The result is a substantial attractive contribution to the free energy, estimated in  $\approx -13$~$k_\mathrm{B}T$ for typical values of $N$.\\
Given $U_\mathrm{hyb}$ in Eq.~\ref{eqn:DNAenergy1}, the term $U_\mathrm{DNA}$ in Eq.~\ref{eqn:energy0} is calculated as
\begin{equation}\label{eqn:DNAenergy4}
U_\mathrm{DNA}(\hat{\theta})=U_\mathrm{hyb}(\hat{\theta})-4Nk_\mathrm{B}T \log\left(\frac{A(\hat{\theta})}{\tilde{A}}\right),
\end{equation}
where the last term is included to account for the changes in DNA confinement following from the stretching of the membranes.\\
Finally, the  $\hat{\theta}$-independent reference energy $U_0$ is calculated for a pair of isolated-unstretched GUVs where only loops can be formed (see Supplementary Methods).\\

Note that the derivation of $U_\mathrm{DNA}(\hat{\theta})$ is based on three assumptions. First, we assume that the separation between the membranes within the adhesion patch  $h$ is equal to the tether length $L$. In the Supplementary Methods we present the generalised theory in which this constraint is relaxed, and demonstrate that, for a fixed geometry, the intervesicle potential exhibits a strong minimum for $h=L$, which justifies the use of the constrained model (Supplementary Fig.~1).\\
Second, we model DNA linkers as freely pivoting rigid rods with point-like sticky ends. This choice is justified in the Supplementary Methods by demonstrating that the deviations caused by assigning a physical size to the sticky end are negligible (see also Supplementary Fig.~2). The validity of the first two assumptions is required to calculate the configurational entropy in Eq.~\ref{eqn:confent}.\\
Third, we neglect steric interactions between linkers, which is justified for sufficiently low DNA coverage. The combinatorial calculation of $U_\mathrm{hyb}$ relies on this assumption. In Supplementary Fig.~3 we show the overall DNA density within ($\rho_\mathrm{DNA}^\mathrm{in}$) and outside ($\rho_\mathrm{DNA}^\mathrm{out}$) the patch. A highest density of $\rho_\mathrm{DNA}^\mathrm{max}\approx 1500$ $\mu$m$^{-2}$ is reached at low temperature within the patch. The ideal gas approximation is justified if $\rho_\mathrm{DNA}^\mathrm{max} B_2 \ll1$, where $B_2$ is the second virial coefficient. We can conservatively estimate $B_2 \lesssim L^2 \approx 1.25 \times 10^{-4}$~$\mu$m$^2$, resulting in $\rho_\mathrm{DNA}^\mathrm{max} B_2 \lesssim 0.18$. Since we used the maximum measured value of the DNA density and a generous overestimation of the second virial coefficient of hard rods, we can conclude that the ideal-tethers approximation is justified.\\
Note also that assuming inter-membrane distance equal to $L$, and neglecting DNA-DNA steric interactions, guarantee a uniform distribution of unbound tethers and loops throughout the entire surface of the membrane, within and outside the patch area. In the Supplementary Methods we quantify the depletion of loops and free tethers from the patch region expected for $h<L$.\\

By combining Eqs.~\ref{eqn:energy1},~\ref{eqn:DNAenergy1}, and \ref{eqn:DNAenergy4} we obtain an analytical expression for the interaction energy between two identical DNA-GUVs. This can be minimised with respect to $\hat{\theta}$, in order to predict the values of all geometrical observables as well as the fraction of formed DNA bonds.\\
Our model contains six input parameters listed in Table~\ref{TableParameters}, which are either measured or estimated from literature data. The area thermal expansion coefficient $\alpha = 1.3 \pm 0.7 \times 10^{-3}$\,K$^{-1}$ is estimated from several vesicles as explained in the Methods section. The density of DNA linkers per unit area is estimated from calibrated fluorescence measurements as  $\rho_\mathrm{DNA}=390\pm90$~linkers$~\mu$m$^{-2}$, including both $a$ and $a'$ tethers. The coating density is then used to compute the overall number of $a/a'$ tethers per vesicle: $N=2\pi R_0^2 \rho_\mathrm{DNA}$. The reference radius $R_0$ is calculated as the average between the radii of each pair of adhering vesicles at $T \approx T_0$, as measured from image analysis. The stretching modulus of DOPC GUVs is estimated from literature data as $K_\mathrm{a}=240\pm90$~mN~m$^{-1}$~\cite{Rawicz_BJ_2000,Rawicz_BJ_2008,Fa_BBA_2007,Pan_BJ_2008}. The conservative errorbar is calculated to cover the entire interval of values reported in literature. The length of the 43 base-pair-long double-stranded spacer $L=14.5$~nm is calculated using the widely accepted estimate of $0.338$~nm~per~base-pair \cite{Smith_Science_1996}. The hybridisation enthalpy and entropy of the sticky ends are evaluated using nearest-neighbours rules~\cite{SantaLucia_PNAS_1998} as $\Delta H^0=-68.5\pm3$~kcal~mol$^{-1}$ and $\Delta S^0=-193.5\pm8$~cal~mol$^{-1}$~K$^{-1}$ respectively. Errorbars are included to correct for possible repulsive effects arising from the presence of {inert tails}~\cite{Di-Michele_JACS_2014}: non-hybridising bases neighbouring the sticky ends. To date, the free energy shift caused by inert tails has only been quantified for the case of single-stranded DNA, while in the present case the inert tails are identified as the double-stranded spacers. For the case of ssDNA tails, the repulsion is found to approximately compensate the stabilising effect of the dangling base: the first non-hybridised base neighbouring a DNA duplex~\cite{Di-Michele_JACS_2014}. In our case the dangling base is the adenine separating the sticky end from the dsDNA spacer (see Fig.~\ref{Figure1}a). The errorbars are calculated such that the minimum and maximium values of $\Delta H^0$ and $\Delta S^0$ correspond to the nearest-neighbour estimates calculated with and without the stabilising effect of the dangling base.\\
The only fitting parameter of our theory is the {neutral temperature} $T_0$, i.e. the temperature at which the vesicles have reduced volume $v=1$ (see Eq.~\ref{eqn:reducedvolume}). $T_0$~is independently fitted for every pair of vesicles tested experimentally. However, before carrying out the measurements, our samples are conditioned with a series of heating/cooling cycles during which  $T_0$ is found to relax to values close to the minimum temperature reached. This relaxation corresponding to a drop in $v$, could be caused either by a reduction of the inside volume or by an increase of the bilayer surface area as explained in the Methods section.\\ 

In Fig.~\ref{Figure2}b-g we compare experimental data with theoretical predictions for the morphological observables. Solid lines indicate the average values, shaded regions indicate uncertainties calculated by propagating the errorbars of the model parameters (Table~\ref{TableParameters}) as explained in the Methods section.  In the inset of Fig.~\ref{Figure2}b we notice that that the predicted contact angle $\theta$ deviates slightly from the unstretched contact angle $\tilde{\theta}$ for $T>T_0$. Higher DNA coverage causes further stretching of the membrane as demonstrated in Supplementary Fig.~4. Quantitative agreement is found for all the morphological observables. Errors for different parameters are independently propagated in order to disentangle their effect on the predictions of the model, and displayed as shaded regions (see Methods section). As expected, the large relative error in $\alpha$ (light blue) substantially contributes to the errorbars of all the morphological observables, in particular at high temperatures, i.e. when $T\gg T_0$. At low temperature, where the DNA-mediated adhesion induces a noticeable stretching of the membrane, the value of the elastic modulus $K_\mathrm{a}$ also plays a role (light green, see zoom in Fig.~\ref{Figure2}b). Uncertainties in the DNA coverage density $\rho_\mathrm{DNA}$, the hybridisation enthalpy $\Delta H^0$, and entropy $\Delta S^0$, play a comparatively minor role in predicting the geometry of the GUV-pairs (pink).\\

In Fig.~\ref{Figure3}e we show the equilibrium values of the fraction of bridges and loops as predicted by the theory. For the case of $x_\mathrm{b}$ we obtain an experimental evaluation by fluorescence intensity measurements (see Fig.~\ref{Figure3}f and Methods). We find a good agreement between theory and experiments, with a slight mismatch at low temperatures. We notice that, similarly to the case of geometrical observables, the uncertainties in the predictions for the fraction of bridges and loops mainly derive from errors in $\alpha$. 
Uncertainties in the hybridisation free energy of the sticky ends and the DNA coating density have little effect on the nature of DNA bonds at low temperature. At high temperature, uncertainties in the DNA degrees of freedom have the clear effect of causing a shift in the melting temperature, that is the temperature at which $x_\mathrm{l}$ and $x_\mathrm{b}$ drop substantially.\\


\section{Discussion}
Error-propagation analysis indicates that, at low temperature, $x_\mathrm{b}$ and $x_\mathrm{l}$ are not very sensitive to the details of the DNA coating, and mostly influenced by the shape of the vesicles (see also Supplementary Fig.~4). The strength of the attractive interactions is in turn also determined by the geometry.
This important remark can be rationalised by combining Eqs.~\ref{eqn:DNAenergy2}, \ref{eqn:DNAenergy3}, and the definition of $\Delta G_\mathrm{b/l}$ (Eqs.~\ref{eqn:singleenergy}-\ref{eqn:transent}), which imply that the equilibrium ratio between the fraction of loops and bridges is uniquely determined  by the relative patch area:  $x_\mathrm{b}/x_\mathrm{l}=A_\mathrm{p}^\mathrm{e}/A$. 
In other words, the formation of intra- or inter-particle bonds is controlled by geometry via the translational entropy. Given that at low temperature $x_\mathrm{b}\simeq 1-x_\mathrm{l}$, the substantial agreement between experimental and predicted $A_\mathrm{p}$, $A$, and $x_\mathrm{b}$, provides an experimental proof of this prediction. This remarkable coupling mechanism could be exploited to design complex interaction schemes in which the competition between loop and bridge formation, and thereby the strength of the adhesive interactions, is controlled by the geometry of the substrates. To date, a similar control over the number of DNA bridges had only been demonstrated in silico through a careful design of complex coating schemes involving 4 or more different linkers with competing interactions \cite{Angioletti-Uberti_NMat_2012,Mognetti_SoftMatter_2012}.\\

We tested the predictions from our model on several pairs of adhering vesicles. The data from 7 pairs of vesicles are summarised in Fig.~\ref{Figure4}, where we plot the relative deviations $(X_\mathrm{experiment}-X_\mathrm{model})/X_\mathrm{model}$ for $X=\theta$, $A_\mathrm{p}$, $D$, and $x_\mathrm{b}$.\\

Because of the simplified treatment of the geometry, our model best reproduces experimental data for pairs of GUVs having similar size and excess area. Large differences in $R_0$ or in the reduced volume lead to significant changes in the geometry of the system, for instance very curved adhesion patches. The framework we present can be extend to arbitrary size and excess area differences but numerical methods would be required to compute the geometry.\\

The effect we described for the case of two vesicles has even more striking consequences in large clusters or networks of DNA-GUVs, as demonstrated in Fig.~\ref{Figure5} and Supplementary Movie 2.
At high temperature (panel a), the high coordination causes the multiple adhesion patches to merge together: the GUV network forms a wet-foam material with no interstices. Upon cooling, the network expands and pores open (panel b). At low temperatures the patches are almost point-like and the expanded network acquires the morphology of a packing of hard spheres (panel c). The morphological changes are qualitatively reversible: upon heating from low temperature the packing recovers the compact morphology (panels d-e). We notice that, especially small vesicles, tend to rearrange their position during thermal cycles. The response we observe is  unique to vesicles and would not occur for stiffer emulsion droplets in which the high interfacial tension prevents substantial deformations, as demonstrated in the Supplementary Methods and Supplementary Fig.~5.\\

%
%
In summary, we observe an unconventional thermal response in pairs and extended networks of DNA-GUVs. The adhesion area between neighbouring vesicles contracts upon cooling, leading to an expansion along the axes of the bonds. This effect arises from the competition between the DNA-induced forces that promote the formation of an adhesive patch area, and the thermal contraction of the lipid bilayers, which tends to restore the spherical shape of the GUVs and reduce the contact area. For the case of pairs of DNA-GUVs of similar size and excess area, we present a theory capable of predicting the temperature dependence of all the geometrical observables as well as the fraction of formed DNA-bridges.
The system we describe represents one of the few experimental examples of a material self-assembled from DNA-mediated interactions that displays structural responsiveness to external stimuli. The simultaneous characterisation of the morphology of the aggregates and the distribution of DNA ligands allows for a direct comparison with the predictions of our theory, which considers both the statistical mechanics of multivalent interactions and the mechanical deformation of very compliant units. The quantitative agreement provides a direct proof of the key role played by entropic confinement, which couples the DNA degrees of freedom to the geometry of the GUVs, and opens the way to new strategies for designing multivalent interactions.
The effect we discovered has striking consequences on the morphology of large clusters of tethered GUVs, that upon cooling, turn from wet-foam structures with no interstices to porous hard-sphere packings with point-like contacts.\\
A material with these features holds great promise in practical applications. A membrane made of tethered vesicles can serve as a tuneable filter with temperature-dependent pore size, useful for size fractionation, sieving, and dialysing mesoscopic objects. Shells of this material will make  capsules capable of releasing their cargo not only upon temperature changes but also when stimulated by osmotic pressure gradients.
The shrinkage of the interstitial volumes observed upon heating can be exploited for confining solutes into geometrically well-defined locations, with possible applications as scaffolds for diffraction studies. Analogously, molecules like cell-cell linking proteins, can be confined in the locations of the point-like patches for biologically relevant studies. We expect the packings of DNA-GUVs to have temperature-dependent rheological properties. In particular, at low temperature, the reduction in the fraction of bridges will cause a drop in the adhesion energy between the GUVs, that should result in lower elastic and viscous moduli. This property could be exploited to build biocompatible scaffolds for tissue regeneration that could be non-invasively injected while kept at low temperature, and then stiffen after thermalising to body temperature. All the materials involved in this study are biocompatible and could be made food grade, which opens up the possibility of exploiting temperature-dependent texture changes in cosmetic and food products. Liposomes can be made readily over a huge size range from 100\,nm to 100\,$\mu$m, which makes all these technologies scalable.


\footnotesize{
\section{Methods}

\subsection{Electroformation}
DOPC GUVs are prepared by standard electroformation in 300mM sucrose solution as explained in the Supplementary Methods~\cite{Angelova_FD_1989,Angelova_PCPS_1992,Estes_2005}.


\subsection{DNA preparation}
The DNA tethers are assembled from two ssDNA strands, one of them (\emph{i}) functionalised with a cholesterol molecule, the second (\emph{ii}) carrying the sticky end:
\begin{enumerate}
\item[\emph{i}] $5'$ -- \textbf{GGA TGG GCA TGC TCT TCC CGT TTT TTA TCA CCC GCC ATA GTA G} \emph{A} [Sticky End] -- $3'$
\item[\emph{ii}] $5'$ -- \textbf{CTA CTA TGG CGG GTG ATA AAA AAC GGG AAG AGC ATG CCC ATC C} \emph{AAAA} [Cholesterol TEG] -- $3'$
\end{enumerate}
The bold letters indicate the segments forming the ds spacer, the italic letters the inert flexible spacers.
The DNA is purchased lyophilised (Integrated DNA Technology), reconstituted in TE buffer (10\,mM tris(hydroxymethyl)aminomethane, 1\,mM ethylenediaminetetra acetic acid, Sigma Aldrich), aliquoted and stored at {-20}$^\circ$C. For assembling the constructs we dilute equal amounts of the two single-strands to 1.6\,$\mu$M in TE buffer with added 100\,mM NaCl. Hybridisation is carried out by ramping down the temperature from 90 to 4$^\circ$C at a rate of $-0.2^\circ$C~min$^{-1}$ using a PCR machine (Eppendorf Mastercycler). We monitored the correct assembly of the structures by repeating the procedure on a UV absorbance spectrophotometer (Supplementary Methods and Supplementary Fig.~6) and tested the assembled constructs with gel electrophoresis (Supplementary Methods and Supplementary Fig.~7).\\


\subsection{Functionalisation and sample preparation}
Functionalisation of the GUVs is carried out by diluting 10\,$\mu$l of electroformed vesicle solution in 90\,$\mu$l of iso-osmolar solution containing TE buffer, 87\,mM glucose, 100\,mM NaCl, 2\,$\mu$M SYTO9 nucleic acid stain (Molecular Probes), and overall 10\,nM DNA constructs, with equal molarity of $a$ and $a'$ strands. The vesicle solution is mixed and incubated for 1h at room temperature to allow grafting. The liposome solutions are then injected in thin-walled glass chambers consisting of a silicon rubber spacer (Altec Products Limited) sandwiched between two microscope coverslips. The  chamber is then sealed using rapid epoxy glue (Araldite). Before usage, coverslips (Menzel-Gl\"{a}ser) are soaked into an alkaline surfactant solution (1\% Hellmanex, Hellma), brought to a boil, sonicated at 90$^\circ$C for 15 minutes, rinsed in double-distilled water, sonicated once more and finally rinsed again~\cite{Di-Michele_NatComm_2013}. Glass is passivated with 1\% bovine serum albumine solution (Sigma Aldrich).\\


\subsection{Temperature cycling and imaging}
The samples are imaged on Nikon Eclipse Ti-E inverted epifluorescence microscope using a Nikon Plan Fluor 40\,$\times$/0.75 dry objective and an Andor iXon3 897 EM-CCD camera. The perfect-focus system (Nikon) allows to keep the sample in focus during thermal drifts. Fluorescence excitation is produced by a blue LED source (Cree XPEBLU, 485\,nm). The temperature of the sample is controlled with an home-made computer-controlled Peltier device. The temperature sensor, a thermocouple, is placed in direct contact with the sample chamber.\\
Before carrying out quantitative measurements we condition freshly prepared samples through a few cooling/heating cycles. This treatment allows the relaxation of the excess area of the vesicles. In Supplementary Fig.~8a we show the temperature dependence of the contact angle of a vesicle in a sample undergoing a thermal cycle for the first time. When cooling down from $40^\circ$C, the contact angle rapidly decreases, indicating a small excess area at high $T$, or, analogously, a large $T_0$. However, as the temperature is further decreased, the pairs exhibit a series of relaxation events, visible as sudden jumps in $\theta$. 
These events are due to drops in the reduced volume of the GUVs, which can be either caused by a drop in the inner volume, or by an increase in the overall membrane area. The latter scenario may be caused by small vesicles and other lipid aggregates merging the large liposomes. Drops in the inner volume may be caused by the opening of transient pores, induced in turn by the increase in membrane tension~\cite{Sandre_PNAS_1999}. In Supplementary Fig.~8b we demonstrate that some of the relaxation events observed in the contact angle indeed correlate with drops in the measured volume of the liposome. The relaxation is irreversible: when the minimum temperature is reached and the samples are heated up again, the vesicles exhibit an increased excess area, corresponding to a lower $T_0$. After one or a small number of preliminary cycles we do not observe further relaxation events and the temperature response of pairs of adhering GUVs becomes fully reversible. The {neutral temperature} is found to relax to values close to the minimum temperature reached during the preliminary cycles.\\
After this conditioning, data are collected while cooling down from $40$ to $\simeq 0^\circ$C at a rate of $\simeq-2^\circ$C min$^{-1}$ and heat back up to $40^\circ$C at the same rate. Epifluorescence snapshots are recorded at $1^\circ$C intervals. 


\subsection{Checking specificity of the bonding}
To check for the specificity of the DNA-mediated interactions we prepare samples in which 100\% tethers carry an $a'$ sticky end and compare them with conventional samples in which both $a$ and $a'$ are present. As shown in Supplementary Fig.~9, samples with a single sticky do not exhibit intervesicle adhesion (panel a), indicating that the attraction found when both $a$ and $a'$ tethers are used (panel b) is ascribable to specific base pairing. No adhesion is observed in tests performed at varying temperature (see Supplementary Movie~3).

\subsection{Image analysis}\label{ImageAnalysis}
Images of adhering vesicles are analysed using custom scripts written in Matlab{\small\circledR} to characterise geometry and ligands distribution. A region of interest (ROI), containing the adhering pair, is selected (Fig.~\ref{Figure2}a). To correct for thermal drift and Brownian motion, the ROI is automatically re-centered at every frame. Each frame is processed with a bandpass filter to remove pixel noise and flatten the background. A straight line fitting the bright adhesion patch is used to segment the image in two halves, each containing one of the GUVs. The adhesion area is masked out from both the images before applying an edge-detection filter to highlight the contour of the vesicles. The filtered images of the two vesicles are fitted with circles $C_1$ and $C_2$, of radii $R_1$ and $R_2$ respectively (blue and red circles in Fig.~\ref{Figure2}$a$). The adhesion patch is fitted with a circle $C_\mathrm{p}$, of radius $R_\mathrm{p}$. The arc $P$ of $C_\mathrm{p}$ corresponding to the adhesion area (green line in Fig.~\ref{Figure2}a) is delimited by the averaged intersection of $C_\mathrm{p}$ with $C_1$ and $C_2$. The area of the adhesion patch is calculated as
\begin{equation}
A_\mathrm{p}=2\pi R_\mathrm{p} \left(R_\mathrm{p}-\sqrt{R_\mathrm{p}^2-L_\mathrm{p}^2/4}\right),
\end{equation}
where $L_\mathrm{p}$ is the length of the chord subtended by the arc $P$.
The overall area of the two vesicles, including the non-adhering portion and the adhesion area, is calculated as
\begin{equation}
A_{1/2}=4\pi R_{1/2}^2 + A_\mathrm{p}- 2\pi R_{1/2} \left(R_{1/2}-\sqrt{R_{1/2}^2-L_\mathrm{p}^2/4}\right).
\end{equation}
Neglecting the curvature of the adhesion patch, the contact angles are given by
\begin{equation}
\theta_{1/2}=\arcsin\left(\frac{L_\mathrm{p}}{2R_{1/2}}\right).
\end{equation}
As shown in Fig.~3f, we define regions of interest around circles  $C_1$ and  $C_2$ and arc $P$. These are used to mask the raw image and calculate the average fluorescence intensity $I_\mathrm{v}$ of the free (non-adhering) portion of the vesicles and the adhesion patch ($I_\mathrm{p}$). For each frame, we measure the average background fluorescence $I_\mathrm{b}$ from an empty region of the raw image. The intercalating dye used to visualise the DNA only fluoresces when bound to dsDNA. In our constructs the largest fraction of the dsDNA makes up the rigid spacers of the ligands. When two ligands bind to each other, the dsDNA content, and therefore the fluorescent intensity, increases by 10\% (two bound/unbound ligands have 95/86 paired bases). By neglecting this effect, we can assume that, regardless of the number of bonds, the DNA surface density is proportional to the fluorescence intensity diminished by the background fluorescence.  For the  non-adhering portion of the membranes and the adhesion patches we find DNA concentrations per unit area $\rho^\mathrm{DNA}_\mathrm{v}\propto I_\mathrm{v} - I_\mathrm{b}$ and $\rho^\mathrm{DNA}_\mathrm{p}\propto I_\mathrm{p} - I_\mathrm{b}$ respectively. By definition, all the bridges are confined to the adhesion area whereas unbound tethers and loops can be located anywhere on the membrane. If we neglect steric interactions between DNA constructs (see discussion below on the ideal gas approximation), we can assume that the number of DNA ligands that are either unbound or involved in a loop is proportional to the fluorescence intensity measured on the non-adhering region integrated over the whole area of the vesicle
\begin{equation}\label{eqn:LoopsUnboundNumImages}
N_\mathrm{l}+N_\mathrm{u} = \rho^\mathrm{DNA}_\mathrm{v} A\propto \left(I_\mathrm{v} - I_\mathrm{b}\right) A.
\end{equation}
Analogously, the number of ligands involved in a bridge can be estimated by integrating the fluorescence intensity measured on the contact area, diminished by the intensity of the non-adhering regions, over the area of the contact patch
\begin{equation}\label{eqn:BridgesNumImages}
N_\mathrm{b} = \frac{1}{2} \left(\rho^\mathrm{DNA}_\mathrm{p} - \rho^\mathrm{DNA}_\mathrm{v}\right)A_\mathrm{p}\propto \frac{1}{2}  \left(I_\mathrm{p} - I_\mathrm{v}\right) A_\mathrm{p}.
\end{equation}
In Eq. \ref{eqn:BridgesNumImages} the factor $1/2$ accounts for the fact that both vesicles contribute to the fluorescence detected in the adhesion region. Using Eq. \ref{eqn:LoopsUnboundNumImages} and \ref{eqn:BridgesNumImages} the fraction of bridges can be estimated as
\begin{equation}\label{eqn:BridgesFracImages}
x_\mathrm{b}=\frac{N_\mathrm{b}}{N_\mathrm{b} + N_\mathrm{l} + N_\mathrm{u}} = \frac{\left(I_\mathrm{p} - I_\mathrm{v}\right) A_\mathrm{p}}{2  \left(I_\mathrm{v} - I_\mathrm{b}\right) A+ \left(I_\mathrm{p} - I_\mathrm{v}\right) A_\mathrm{p}}.
\end{equation}
An expression for the fraction of loops $x_\mathrm{l}$ analogous to Eq. \ref{eqn:BridgesFracImages} is not possible because in our system we cannot distinguish between unbound ligands and ligands involved in a loop.
Note that the intensity of SYTO9 stain decreases monotonically with temperature. The fact that in Eq. \ref{eqn:BridgesFracImages} we calculate a ratio between intensities corrects for this effect. Bleaching is not observed over the duration of the experiments (several hours).\\


\subsection{Estimating DNA coverage}
We evaluate the DNA coverage density from confocal images acquired using a  Leica TCS SP5 microscope equipped with an HCX PL APO CS 100/1.4 oil immersion objective and an Ar-ion laser line (488 nm) as the excitation source.\\ We produce samples using 100\% of $a'$ tethers: in the absence of complementary linkers vesicles do not adhere to each other. We reconstruct the contour of large GUVs from equatorial cross sections. The average fluorescence intensity $I_\mathrm{v}$ is calculated within the green shaded region in Supplementary Fig.~10a. The fluorescent intensity due to DNA per unit surface is given by
\begin{equation}\label{eqn:DNACov1}
J=\frac{I_\mathrm{v}-I_\mathrm{d}}{2 \pi R \delta},
\end{equation}
where $R$ is the radius of the vesicle and $\delta$ the $z$-resolution of the confocal microscope. In Eq. \ref{eqn:DNACov1} $I_\mathrm{v}$, is diminished by the dark signal of the sensor $I_\mathrm{d}$, measured in the absence of laser light. We can convert $J$ to surface coverage by measuring the fluorescent intensity from a set of samples with a known number density $n$ of free DNA, in the absence of GUVs. As demonstrated in Supplementary Fig.~10b, the reference intensity, dimished by the dark signal, scales linearly with $n$
\begin{equation}\label{eqn:DNACov2}
I_\mathrm{ref}(n)-I_\mathrm{d} \simeq C d^2 n.
\end{equation}
In Eq. \ref{eqn:DNACov2} $d=0.076$ $\mu$m is the pixel size of the camera and we fit $C=2.7$ $\mu$m.
The DNA coverage per unit of bilayer surface can be readily calculated as $\rho_\mathrm{DNA}=J\delta/C$. We obtain $\rho_\mathrm{DNA}=390 \pm 90$ strands $\mu$m$^{-2}$ by averaging over 25 vesicles. Note that the slice thickness $\delta$ cancels out from the calculation.\\

\subsection{Estimation of the area thermal expansion coefficient}
We estimate the area thermal expansion coefficient $\alpha$ from image analysis of pairs of adhering vesicles prepared according to the protocol described above. For temperatures sufficiently higher than $T_0$ the DNA-induced stretching is expected to be negligible in comparison with the thermal expansion. On the other hand the adhesion is sufficient to suppress thermal fluctuations, allowing for an accurate estimate of $A(T)$.
For temperatures $20\le T \le 40^\circ$C, $A(T)$ is linearly fitted using Matlab{\small\circledR} to extract $\alpha=\frac{1}{A}\frac{\partial A}{\partial T}$. By averaging over 18 vesicles (9 pairs) we find $\alpha=1.3\pm0.7\times10^{-3}$ K$^{-1}$.

\subsection{Zero-stretching adhesion}
In the presence of an excess area ($v<1$), any adhesive force large enough to overcome thermal fluctuations will cause a deformation of the two vesicles and the formation of an adhesion patch. Given a reduced volume $v$, the zero-stretching contact angle $\tilde{\theta}$ can be derived from simple geometry assuming constant area $A=\tilde{A}$. In particular using
Eqs.\ \ref{eqn:A} and \ref{eqn:V} in Eq.\ \ref{eqn:reducedvolume} we find that $\tilde{\theta}$ is solution of~\cite{Tordeux_PRE_2002}
\begin{equation}\label{eq:zs}
\left(v^2+4\right) \cos^3 \tilde{\theta} - \left(9 v^2 + 12\right) \cos^2 \tilde{\theta} + 27 v^2 \cos \tilde{\theta} -27 v^2 + 16 = 0.
\end{equation}

\subsection{Contact-angle dependence of areas and volume}
To minimise the interaction energy in Eq. \ref{eqn:energy1}  an explicit relation between $A$, $A_\mathrm{p}$ and the contact angle $\hat{\theta}$ is needed. From simple geometry we obtain \cite{Ramachandran_Langmuir_2010}
\begin{eqnarray}
&A_\mathrm{p}=\pi R^2(\hat{\theta}) \sin^2 \hat{\theta} \label{eqn:Ap} \\
&A=\pi R^2\left(\hat{\theta} \right) \left(1+\cos \hat{\theta}\right)\left(3-\cos \hat{\theta}\right)  \label{eqn:A} \\
&V=\frac{\pi R^3}{3}\left(1+\cos \hat{\theta}\right)^2\left(2-\cos \hat{\theta} \right)  \label{eqn:V}.
\end{eqnarray}
To make the $\hat{\theta}$-dependence of these observables explicit, we need a further relation between $R$ and $\hat{\theta}$. This can be done under the assumption of constant volume, corresponding of water-impermeable vesicles. By substituting $V=V_0=4/3\pi R_0^3$ in Eq. \ref{eqn:V} we obtain
\begin{equation}\label{eqn:RofTheta}
R=R_0\left[\frac{4}{\left(1+\cos \hat{\theta}\right)^2\left(2-\cos \hat{\theta}\right)}\right]^{1/3},
\end{equation}
which can be inserted into Eqs. \ref{eqn:Ap} and \ref{eqn:A}. Given the evidence of small volume changes in our vesicles, we decide to work under the constant-volume assumption. However, one can find an expression analogous to Eq.~\ref{eqn:RofTheta} for the case of {osmotically equilibrated} vesicles, corresponding to water-permeable, solute-impermeable, membranes. As discussed in the Supplementary Methods, in the regime relevant to our experiments the two assumptions lead to very similar results (see Supplementary Figs 11 and 12).\\

\subsection{Configurational entropy }

For DNA linkers modelled as freely pivoting rigid rods with fixed tethering points placed at distance $d_t$ (Fig.~\ref{Figure2}a), an expression for  the rotational entropy can be derived from Eq.\ B5 of Ref.\ 
\cite{Mognetti_SoftMatter_2012}
\begin{eqnarray}
\exp[\Delta S^\mathrm{rot}_\mathrm{b/l}/k_\mathrm{B}] &=&
{ f^\mathrm{cut}_\mathrm{l/b} \over 2\pi \rho_0 L^2 } { 1 \over d_t }  \, ,
\label{eqn:info_conf1}
\end{eqnarray}
where $f^\mathrm{cut}_\mathrm{b}=1$ for bridges and $f^\mathrm{cut}_\mathrm{l}=1/2$ for loops as due to the 
fact that in the latter case the orbit of the hybridised rods is half-excluded by the 
(flat) surface of the GUV. The rotational entropy in Eq. \ref{eqn:rotent} can be obtained by averaging over all the possible values of $d_t$
\begin{eqnarray}
\Delta S^\mathrm{rot} &=& k_\mathrm{B} \log\left[  \frac{1}{4\pi \rho_0 L^4} \int_{d_t\leq 2L} \frac{f^\mathrm{cut}_\mathrm{b/l} y}{d_t} \mathrm{d}y \right]\nonumber \\
&=& k_\mathrm{B} \log\left[  \frac{1}{4\pi \rho_0 L^3} \right] \label{eqn:confaverage}.
\end{eqnarray}
In Eq. \ref{eqn:confaverage}, $y$ is the lateral displacement between tethering points (Fig.~\ref{Figure2}a). For loops $d_t=y$, for bridges, assuming a distance $L$ between opposite membranes, $d_t=\sqrt{y^2+L^2}$. The integral is 
bound by the maximal lateral displacement for which $d_t=2L$. Note that after averaging there is not difference in rotational entropy between loops and bridges.
 For mobile constructs 
we need to account for the loss in translational entropy (Eq.\ \ref{eqn:transent}).
This is obtained by including a factor $4\pi L^2 \hat A_\mathrm{b/l}(\hat \theta)/
A^2(\theta)$ that is the ratio between all positions available to bound constructs ($4\pi L^2 \hat A_\mathrm{b/l}$) and all of those available to two unbound tethers ($A^2$). These terms result in the configurational entropy given in Eq.\ \ref{eqn:confent}. In this derivation we model the sticky ends as point particles, as justified in the Supplementary Methods. The rigidity of the double-stranded spacer is guaranteed by the fact that the persistence length of dsDNA is $\gtrsim 3 L$ ~\cite{Smith_Science_1996}.\\


\subsection{Effective patch-area for bridge formation}
Note that for the case of bridges, in Eq.~\ref{eqn:transent} we use an effective adhesion area $A_\mathrm{p}^\mathrm{e}>A_\mathrm{p}$. This correction is made to account for the fact that tethers of length $L$ can form bridges not only within the flat contact area but also within a rim surrounding it. With reference to Fig.~\ref{Figure3}b we find $A_\mathrm{p}^\mathrm{e}={A}_\mathrm{p} + 2\pi R L$. The use of the uncorrected $A_\mathrm{p}$ in Eq.~\ref{eqn:transent} would lead to a divergence in $\Delta S^\mathrm{trans}_\mathrm{b}$ due to the fact that $A_\mathrm{p} \rightarrow 0$ for $\hat{\theta} \rightarrow 0$. This artefact would lead to an overestimation of the coupling between the hybridisation free energy for bridges formation and the patch area at low temperatures.\\

\subsection{Derivation of the hybridisation free-energy}
If we neglect steric interactions between the tethers, for given $\Delta G_\mathrm{b/l}$ we can express $U_\mathrm{hyb}$ by taking into account combinatorics. We indicate as $N_1^\mathrm{l}$, $N_2^\mathrm{l}$, $N_1^\mathrm{b}$, $N_2^\mathrm{b}$ the number of loops/bridges formed on vesicles 1 and 2. The free energy of the system is given by 
\begin{widetext}
\begin{equation}\label{eqn:FeeDef}
\exp \left[ -\beta U_\mathrm{hyb} \right] = \sum \Omega \left(N_1^\mathrm{l},N_2^\mathrm{l},N_1^\mathrm{b},N_2^\mathrm{b}\right) \exp \left[ -\beta \left(N_1^\mathrm{l} + N_2^\mathrm{l}\right) \Delta G_\mathrm{l} -\beta \left(N_1^\mathrm{b} + N_2^\mathrm{b}\right) \Delta G_\mathrm{b}   \right],
\end{equation}
\end{widetext}
where $\Omega \left(N_1^\mathrm{l},N_2^\mathrm{l},N_1^\mathrm{b},N_2^\mathrm{b}\right)$ indicates the number of possible configurations with $N_1^\mathrm{l}$ (or $N_2^\mathrm{l}$) loops and $N_1^\mathrm{b}+N_2^\mathrm{b}$ bridges
\begin{widetext}
\begin{equation}
\Omega \left(N_1^\mathrm{l},N_2^\mathrm{l},N_1^\mathrm{b},N_2^\mathrm{b}\right)={N \choose N_1^\mathrm{l}}^2 N_1^\mathrm{l}! {N \choose N_2^\mathrm{l}}^2 N_2^\mathrm{l}!
{N - N_1^\mathrm{l} \choose N_1^\mathrm{b}} {N - N_2^\mathrm{l} \choose N_1^\mathrm{b}} N_1^\mathrm{b}!
{N - N_1^\mathrm{l} \choose N_2^\mathrm{b}} {N - N_2^\mathrm{l} \choose N_2^\mathrm{b}} N_2^\mathrm{b}!.
\end{equation}
\end{widetext}
The sum in Eq. \ref{eqn:FeeDef} is extended to all possible values of $N_{1/2}^\mathrm{l/b}$.\\
By defining the fraction of loops/bridges as $x_{1/2}^\mathrm{l/b} = N_{1/2}^\mathrm{l/b}/N$ and using the Stirling approximation we find
\begin{equation}
\exp\left[ - \beta U_\mathrm{hyb} \right] = \sum \exp\left[ - N A\left(x_{1}^\mathrm{l} ,x_{1}^\mathrm{b} ,x_{2}^\mathrm{l} ,x_{2}^\mathrm{b} \right) \right] \label{eqn:DNAFreeEnergy1},
\end{equation}
with
\begin{widetext}
\begin{align}
A\left(x_{1/2}^\mathrm{l/b} \right) = & x_{1}^\mathrm{l} \log x_{1}^\mathrm{l} + x_{1}^\mathrm{b} \log x_{1}^\mathrm{b} + x_{2}^\mathrm{l} \log x_{2}^\mathrm{l} + x_{2}^\mathrm{b} \log x_{2}^\mathrm{b}+ \nonumber \\
& \left(1 - x_{1}^\mathrm{l} - x_{1}^\mathrm{b}\right) \log \left(1 - x_{1}^\mathrm{l} - x_{1}^\mathrm{b}\right) + \left(1 - x_{2}^\mathrm{l} - x_{2}^\mathrm{b}\right) \log \left(1 - x_{2}^\mathrm{l} - x_{2}^\mathrm{b}\right) +  \nonumber \\
& \left(1 - x_{1}^\mathrm{l} - x_{2}^\mathrm{b}\right) \log \left(1 - x_{1}^\mathrm{l} - x_{2}^\mathrm{b}\right) + \left(1 - x_{2}^\mathrm{l} - x_{1}^\mathrm{b}\right) \log \left(1 - x_{2}^\mathrm{l} - x_{1}^\mathrm{b}\right) +  \nonumber \\
& \left( x_{1}^\mathrm{l} + x_{2}^\mathrm{l} \right) \left(  \beta \Delta G_\mathrm{l}^* +1 \right) +   \left( x_{1}^\mathrm{b} + x_{2}^\mathrm{b} \right) \left(  \beta \Delta G_\mathrm{b}^* +1 \right),
\end{align}
\end{widetext}
Where we define  $\Delta G_\mathrm{l/b}^* =  \Delta G_\mathrm{l/b} -k_\mathrm{B}T\log N$.\\
In the saddle point approximation $N \rightarrow \infty$, the equilibrium fractions of hybridised strands $ \bar{x}_{1/2}^\mathrm{l/b}$ are given by the gap equations
\begin{equation}\label{eqn:Saddle}
\frac{\mathrm{d}}{\mathrm{d} {x}_{1/2}^\mathrm{l/b}} A\left(\bar{x}_{1/2}^\mathrm{l/b}\right) = 0.
\end{equation}
In this limit the sum in Eq.  \ref{eqn:FeeDef} is dominated by the saddle-point
\begin{equation}\label{eqn:DNAFreeEnergy2}
U_\mathrm{hyb}=k_\mathrm{B} T N A\left(\bar{x}_{1/2}^\mathrm{l/b} \right).
\end{equation}
The saddle-point equations in Eq. \ref{eqn:Saddle} are identical to the self-consistent relations derived by Varilly \emph{et al.}~\cite{Varilly_JChemPhys_2012}  and can be solved analytically with the formula derived by Angioletti-Uberti  \emph{et al.}~\cite{Angioletti-Uberti_JCP_2013}. If we look for symmetric solutions with $\bar{x}_\mathrm{l/b}= \bar{x}_{1}^\mathrm{l/b}=\bar{x}_{2}^\mathrm{l/b}$ Eq. \ref{eqn:Saddle} reduces to
\begin{align}
& \frac{\bar{x}_\mathrm{l}}{\left(1-\bar{x}_\mathrm{l}-\bar{x}_\mathrm{b}\right)^2}=\exp \left(-\beta \Delta G_\mathrm{l}^*\right)=q_\mathrm{l} \\
& \frac{\bar{x}_\mathrm{b}}{\left(1-\bar{x}_\mathrm{l}-\bar{x}_\mathrm{b}\right)^2}=\exp \left(-\beta \Delta G_\mathrm{b}^*\right) = q_\mathrm{b}
\end{align}
which can be solved to obtain Eq.~\ref{eqn:DNAenergy2}.
The equilibrium hybridisation free energy in Eq.~\ref{eqn:DNAenergy1} can be then calculated by substituting Eq.~\ref{eqn:DNAenergy2} in Eq. \ref{eqn:DNAFreeEnergy2}.
For simplicity, in the main text, we indicate the equilibrium fractions of hybridised strands as $x_\mathrm{b/l}$, omitting the overlines.


\subsection{Error propagation}
Uncertainties in the model parameters are numerically propagated to extract errorbars on the model predictions. We sample the predictions of the model obtained using input parameters $X=\alpha$, $K_\mathrm{a}$, $\rho_\mathrm{DNA}$, $\Delta H^0$, and $\Delta S^0$ extracted from normal distributions with mean $X_0$ and standard deviation $\Delta X$, taken respectively as the central values ad the errorbars listed in Table~\ref{TableParameters}. Note that, since $\Delta H^0$ and $\Delta S^0$ are naturally coupled, they are sampled using the same normal-random numbers. If independently propagated, variations in $\Delta H^0$ and $\Delta S^0$ would lead to unphysical fluctuations in the melting temperature of the sticky ends.\\
For all the observables shown in Figs.~\ref{Figure2}, and \ref{Figure3}, averages are calculated as the median of the sampled distributions (solid lines). Errobars cover the interval between the 16$^\mathrm{th}$ and the 84$^\mathrm{th}$ percentile, that is 68$\%$ of the sampled points, equal to the fraction of input parameters extracted between $X_0-\Delta X$ and $X_0+\Delta X$.
Light blue bands in Figs.~\ref{Figure2}b-g and \ref{Figure3} are calculated by propagating errors in $\alpha$. Wider bands in light green are calculated by propagating errors in $\alpha$ and $K_\mathrm{a}$. The widest bands in pink and grey bands in Fig.~\ref{Figure4} are calculated by propagating uncertainties in $\alpha$, $K_\mathrm{a}$, $\Delta H^0$, $\Delta S^0$, and $\rho_\mathrm{DNA}$.
For each temperature, the size of the statistical sample is between 1 and 5$\times 10^4$.
}


\section{Acknowledgments}
Funding was provided by  the Ernest Oppenheimer Fund and Emmanuel College Cambridge (LDM), EPSRC Programme Grant CAPITALS number EP/J017566/1 (LP, JK, PC, and LDM) and the Winton Fund for Physics of Sustainability (EE).
\section{Authors Contributions}
LP, EE, PC and LDM designed research; LP, and LDM performed experiments; JK optimised the imaging setup; LP, BM, PC and LDM developed the model;  LP, BM, PC and LDM analysed the data; LP, BM, PC and LDM wrote the paper. All the authors discussed the results and commented on the manuscript.
{\section{Competing Financial Interests}
The authors declare no competing financial interests.}
\clearpage

\begin{figure*}[ht!]
\includegraphics[width=18cm]{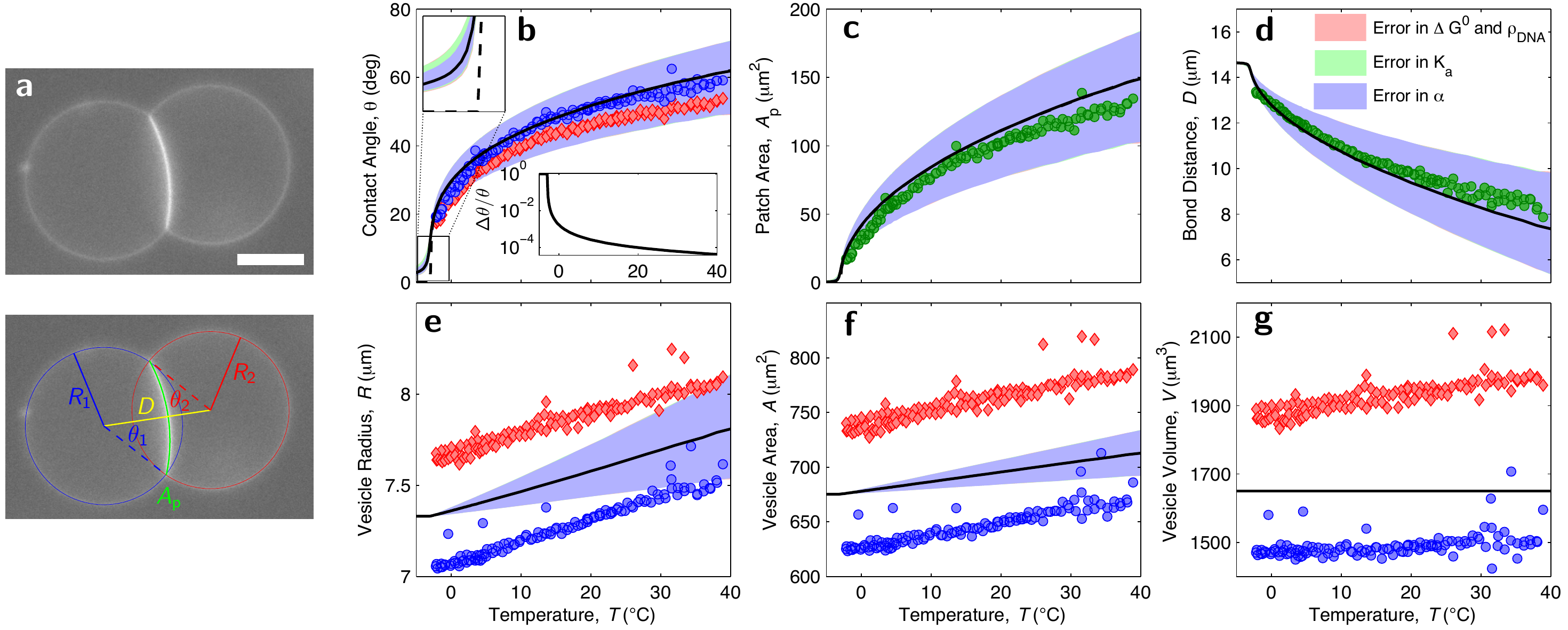}%
\caption{\label{Figure2} \textbf{Temperature dependence of geometrical observables.} \textbf{a}~Snapshot of a pair of adhering DNA-GUVs. Scale bar, 10~$\mu$m. In the bottom image we highlight geometrical features as extracted by our analysis software (see Methods section).  Blue and red solid lines mark the radii $R_{1/2}$ of the spherical segments of the two GUVs. The slightly curved adhesion patch is marked by a green solid line. The contact angles $\theta_{1/2}$ are measured between the segment, of length $D$ (yellow solid line), connecting the centres of the two vesicles, and a radius subtending the patch (red and blue dashed lines). In panels \textbf{b}-\textbf{g} we show the temperature dependence of the geometrical observables: equilibrium contact angle $\theta$~(\textbf{b}), adhesion patch area $A_\mathrm{p}$~(\textbf{c}), centre-to-centre distance $D$~(\textbf{d}), vesicle radius $R$~(\textbf{e}), vesicle total area $A$~(\textbf{f}), and vesicle volume $V$~(\textbf{g}) for a typical pair of adhering vesicles. Symbols indicate experimental data. Red lozenges and blue circles are used to distinguish between the two GUVs. Black solid lines indicate the model predictions calculated using the parameters in Table~\ref{TableParameters} and the fitting parameter $T_0=-3$\,$^\circ$C. Shaded regions indicate the model errorbars. We independently propagate the uncertainties deriving from $\alpha$ (light blue), $K_\mathrm{a}$ (light green), and combine those deriving from $\Delta H^0$, $\Delta S^0$ and $\rho_{DNA}$ (pink). The legend in panel \textbf{d} applies to panels \textbf{b}-\textbf{g}. In panel \textbf{b}, the dashed line (better visible in the zoom, top left) indicates the unstretched contact angle $\tilde{\theta}$. In the inset at the bottom we show the relative deviation between the predicted $\theta$ and the unstretched $\tilde{\theta}$ contact angles: $\Delta \theta / \theta = \left(\theta-\tilde{\theta}\right)/\theta$.}
\end{figure*}

\begin{figure}[ht!]
\includegraphics[width=8.7cm]{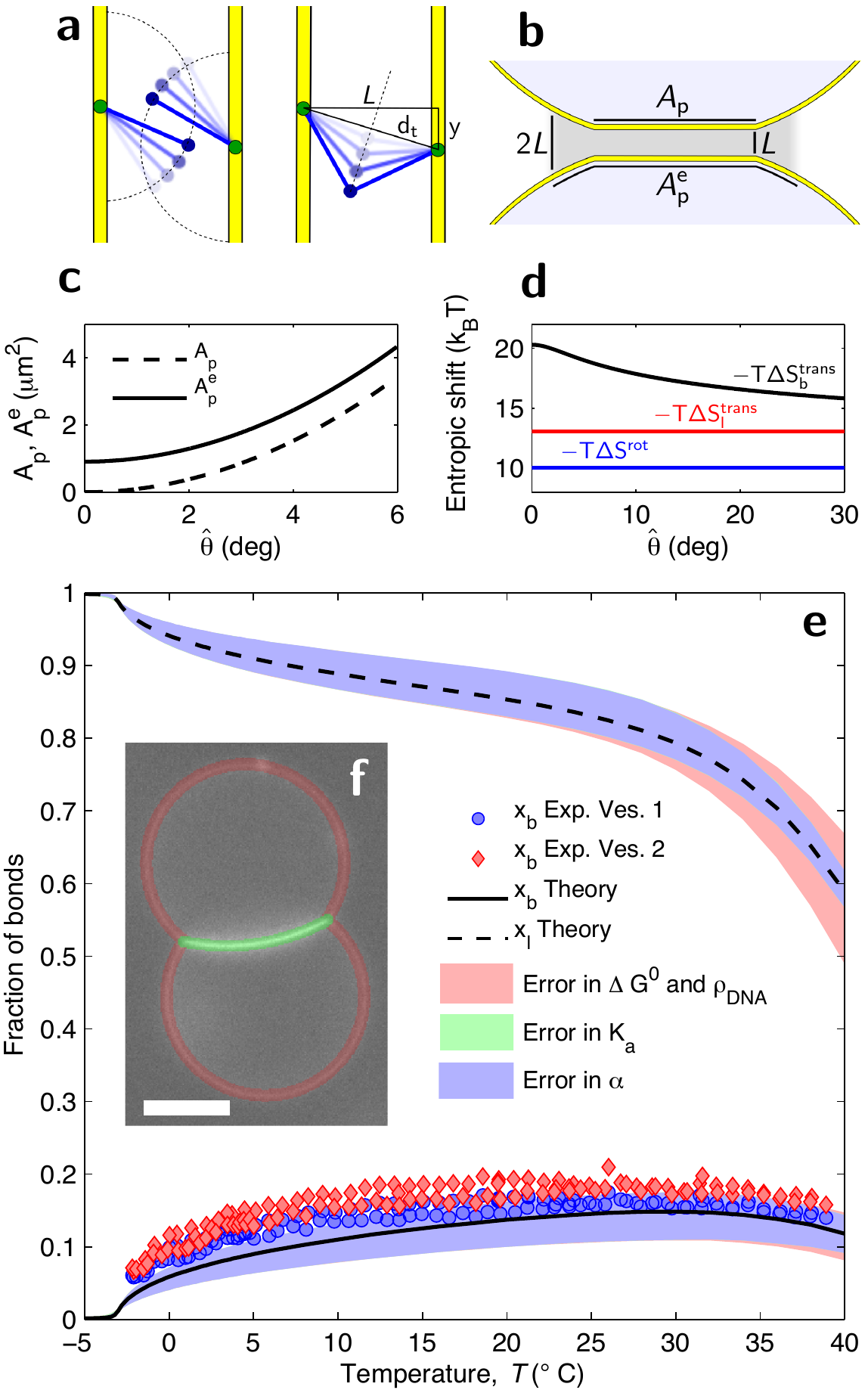}%
\caption{\label{Figure3} \textbf{DNA hybridisation.} \textbf{a}~Sketch demonstrating the loss of orientational entropy following the formation of a bridge. \textbf{b}~Sketch of the effective patch area for bridge formation $A_\mathrm{p}^\mathrm{e}$. \textbf{c}~Comparison between the contact-angle dependence of $A_\mathrm{p}^\mathrm{e}$ and $A_\mathrm{p}$: the latter decays to zero for $\hat{\theta} \rightarrow 0$.  \textbf{d}~Translational entropy loss for bridge (black) and loop (red) formation (Eq.~\ref{eqn:transent}), and rotational entropy loss (blue, Eq.~\ref{eqn:rotent}). \textbf{e}~Fraction of bound tethers: Experimental estimates of $x_\mathrm{b}$ (symbols) and theoretical prediction for the fraction of bridges ($x_\mathrm{b}$, solid line) and loops ($x_\mathrm{l}$, dashed line). Shaded regions indicate the errorbars of the model propagated independently from uncertainties in $\alpha$ (light blue), $K_\mathrm{a}$ (light green), and combining those from $\Delta H^0$, $\Delta S^0$ and $\rho_{DNA}$ (pink). The model parameters are given in Table~\ref{TableParameters}. \textbf{f} Snapshot of a pair of adhering DNA-GUVs. The shaded regions are used to estimate the average fluorescent intensity of the adhering (green) and non-adhering (red) portions of the GUVs. Scale bar, 10~$\mu$m.}
\end{figure}

\begin{figure}[ht!]
\includegraphics[width=8.7cm]{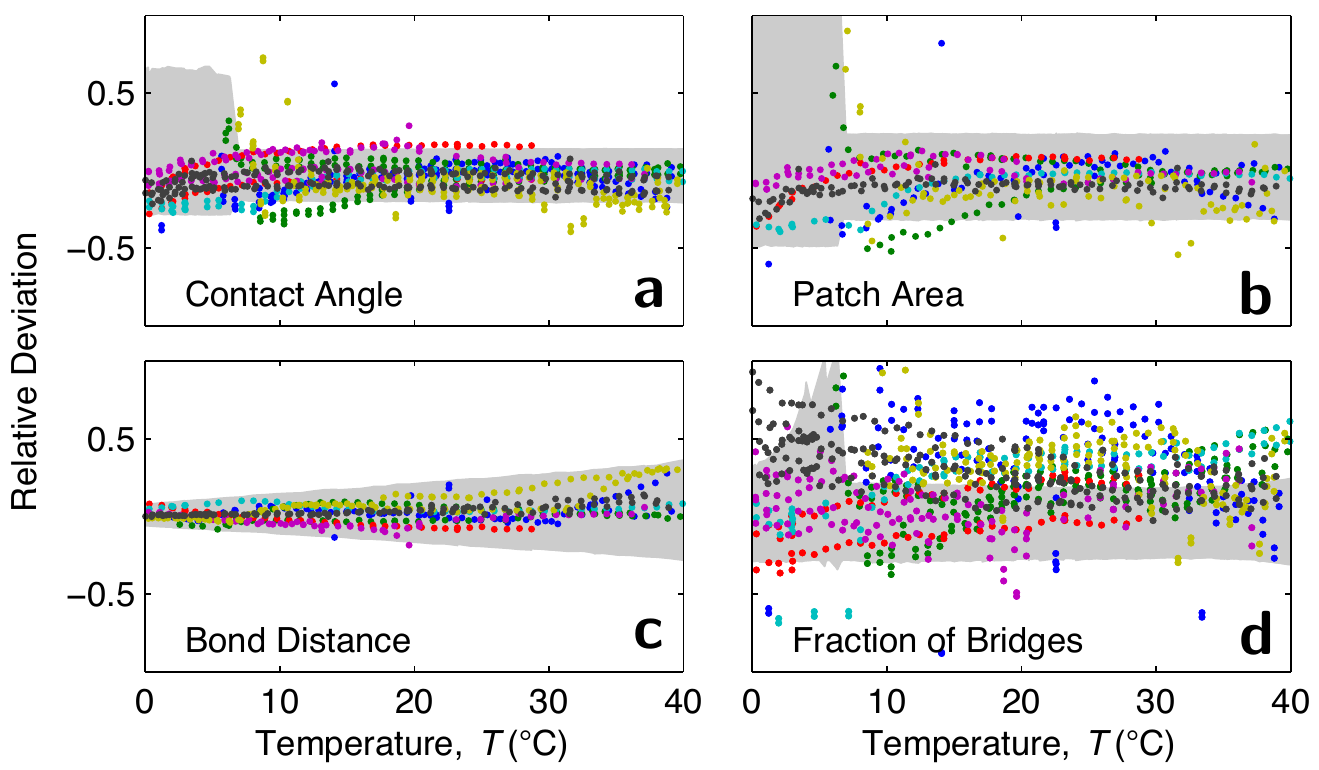}%
\caption{\label{Figure4} { \textbf{Deviation of experimental results from theoretical predictions}. Graphs illustrate the relative deviations  $(X_\mathrm{experiment}-X_\mathrm{model})/X_\mathrm{model}$ for $X$ equal to the contact angle $\theta$ (\textbf{a}), the adhesion-patch area $A_\mathrm{p}$ (\textbf{b}), the bond distance $D$ (\textbf{c}), and the fraction of formed bridges $x_\mathrm{b}$ (\textbf{d}). The results are robust over all 7 pairs of vesicles we have quantified. Different colours indicate different datasets. Grey shaded regions indicate the overall errorbars of the model.}}
\end{figure}

\begin{figure*}[ht!]
\includegraphics[width=18cm]{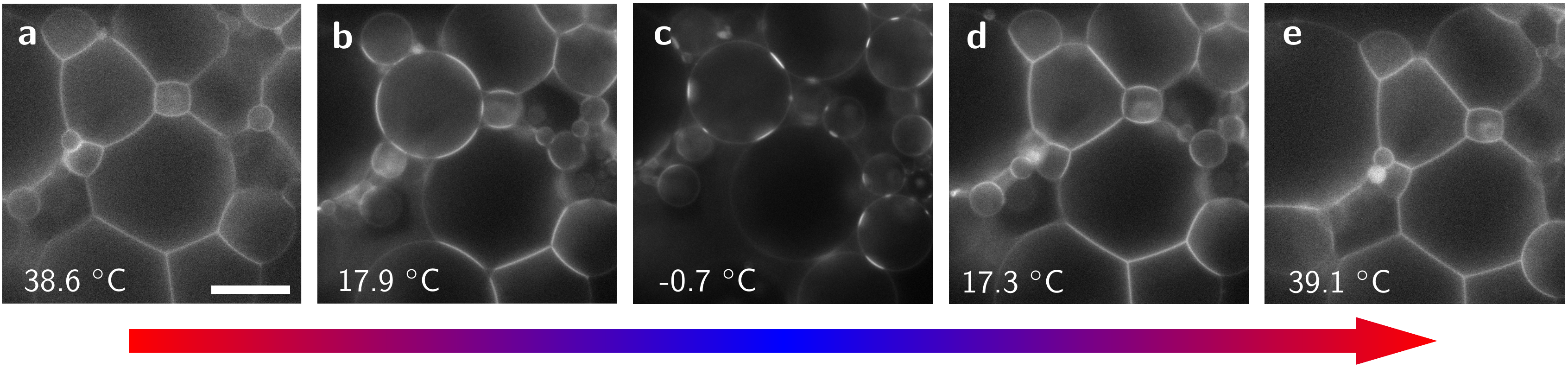}%
\caption{\label{Figure5} \textbf{Thermal response of an extended cluster of linked GUVs showing reversible changes in porosity.} {Snapshots of a network of DNA-GUVs taken upon cooling from high to low temperature (\textbf{a}-\textbf{c}), and subsequently heating up from low to high temperatures (\textbf{c}-\textbf{e}). Scale bar, 20~$\mu$m. For the full sequence see {Supplementary Movie} 2.}}
\end{figure*}
\clearpage

 \begin{table}[ht!]
\caption{\label{TableParameters} \textbf{Input parameters of the model,} estimated experimentally or from literature data. Briefly: The area thermal expansion $\alpha$ and the DNA coating density $\rho_0$ are measured experimentally (see Methods section); the stretching modulus $K_\mathrm{a}$, the length of the double-stranded DNA spacer $L$ and the hybridisation enthalpy/entropy $\Delta H^0$/$\Delta S^0$, are estimated from literature data as explained in the text. The only fitting parameter of the model is the reference temperature $T_0$ (Eq.~\ref{eqn:unstretched}).}
\begin{ruledtabular}
\begin{tabular}{c}
$\alpha = 1.3\pm0.7$ K$^{-1}$\\
$\rho_\mathrm{DNA} = 390\pm90$ $\mu$m$^{-2}$\\
$K_\mathrm{a} = 240\pm90$ mN~m$^{-1}$\\
$L=14.5$ nm\\
$\Delta H^0=-68.5\pm3$ kcal~mol$^{-1}$\\
$\Delta S^0=-193.5\pm8$ cal~mol$^{-1}$~K$^{-1}$\\
\end{tabular}
\end{ruledtabular}
\end{table}

\end{document}